\documentclass[preprint,superscript,address,showkeys]{revtex4}
\usepackage{amsmath,amssymb}
\usepackage{epsfig}
\usepackage{graphics}
\usepackage{graphicx}
\usepackage{float}
\usepackage{dcolumn}

\begin{document}
\title{VIBRATIONAL AND GHOST-VIBRATIONAL RESONANCES IN A MODIFIED CHUA'S CIRCUIT MODEL EQUATION}
\author{K.~Abirami}
\email{abirami@cnld.bdu.ac.in}
\author{S.~Rajasekar}
\email{rajasekar@cnld.bdu.ac.in}
\affiliation{School of Physics, Bharathidasan University, Tiruchirapalli 620 024, Tamilnadu, India}
\author{M.A.F.~Sanjuan}
\email{miguel.sanjuan@urjc.es}
\affiliation{Departamento de F\'{\i}sica,  Universidad Rey Juan Carlos, Tulip\'an s/n, 28933 M\'ostoles,  Madrid, Spain}

\begin{abstract}
The role of the number of breakpoints $N$ in the sawtooth form of the characteristic function in the modified Chua's circuit model equation on vibrational and ghost-vibrational resonances is investigated in this paper. To observe vibrational resonance the system should be driven by two periodic forces of frequencies $\omega$ and $\Omega$, with $\Omega\gg\omega$. Resonance occurs at the frequency $\omega$ when the amplitude of the high-frequency force is varied. When the system is subjected to an input signal containing multi-frequencies which are higher-order of a certain (missing) fundamental frequency, then a resonance at the missing fundamental frequency is induced by the high-frequency input signal and is called ghost-vibrational resonance. In both types of resonances, the number of resonances is $N$ and hysteresis occurs in each resonance region. There are some similarities and differences in these two resonance phenomena. We report in detail the influence of the role of number of breakpoints $N$ on the features of vibrational and ghost-vibrational resonances.
\end{abstract}
\maketitle

\section{Introduction}
\label{int}
Chua's circuit is the most widely investigated nonlinear circuit. The voltage-current characteristic of the operational amplifier based Chua's diode, the nonlinear element in the Chua's circuit, is piecewise linear. The Chua's circuit is found to display a rich variety of nonlinear dynamics. Over the past three decades the occurrence of different kinds of nonlinear phenomena in this circuit has been investigated experimentally, theoretically and numerically [Fortuna {\emph{et al.}},  2009; Kilic, 2010; Madan, 1993]. An interesting attractor of the Chua's circuit is the double-scroll orbit. The original Chua's circuit is modified to generate multi-scroll orbits by replacing the three-segment piecewise linear function with piecewise linear functions having multiple breakpoints. The dynamics of a modified Chua's circuit with multiple piecewise linear segments [Alaoui, 1999; Suykens \& Vandewalle, 1993; Zhong {\emph{et al.}}], sigmoid function [Mahla \& Badan Palhares, 1993], a discontinuous function [Khibnik {\emph{et al.}}, 1993; Lamarque {\emph{et al.}}, 1999], sine function [Xia {\emph{et al.}}, 2003], sign function [Yal\c{c}in {\emph{et al.}}, 2001], hyperbolic tangent function [\"{O}zo\v{g}uz {\emph{et al.}}, 2002], nonlinear term $x|x|$ [Xia {\emph{et al.}}, 2003], saturated function [L\"{u} {\emph{et al.}}, 2004] and a sawtooth function [Yu {\emph{et al.}}, 2007] have been reported. A systematic methodology for designing circuits to generate $n$-scroll orbits has been developed [Zhong {\emph{et al.}}, 2002; Yu, {\emph{et al.}}, 2005; Lu \& Chen, 2006; Campos-Canton {\emph{et al.}}, 2010]. 

In recent years, the study of the $n$-scroll Chua's circuit has received a great deal of interest. For example, adaptive control of chaotic dynamics [Zou {\emph{et al.}}, 2006], existence of $n$-scroll chaotic attractors, global stability of equilibrium points and feedback control laws for synchronization [Xu {\emph{et al.}}, 2009], stabilization and synchronization of $n$-scroll chaotic orbits [Xu \& Yu, 2009], controlling of unstable equilibrium points and periodic orbits [Boukabou {\emph{et al.}}, 2009], diffusion dynamics and characteristic features of first passage times to $n$th scroll attractor and residence times on a scroll attractor [Sakthivel {\emph{et al.}}, 2012] were analysed. Furthermore, the influence of the number of equilibrium points on the characteristics of stochastic and coherence resonances [Arathi {\emph{et al.}}, 2013] have been investigated. The goal of the present paper is to investigate the resonance dynamics induced by a high-frequency periodic force in the presence of single and multiple low-frequency periodic forces in the modified Chua's circuit model equation.

In a nonlinear system driven by a biharmonic force with two frequencies $\omega$ and $\Omega$, with $\Omega\gg\omega$, when the amplitude $g$ of the high-frequency force is varied, the response amplitude at the low-frequency $\omega$ exhibits a resonance. This high-frequency force induced resonance is called \textit{vibrational resonance} [Landa \& McClintock, 2000; Blekhman \& Landa, 2004]. The occurrence of vibrational resonance has been studied in monostable [Jeyakumari {\emph{et al.}}, 2009], bistable [Landa \& McClintock, 2000; Blekhman \& Landa, 2004], multistable [Rajasekar {\emph{et al.}}, 2011], excitable [Ullner {\emph{et al.}}, 2003] and small-world networks [Deng {\emph{et al.}}, 2010]. Experimental evidence of vibrational resonance in an excitable circuit [Ullner {\emph{et al.}}, 2003] and in a bistable vertical cavity surface emitting laser [Chizhevsky \& Giacomelli, 2008] have been reported. When an excitable system is driven by an input signal containing multi-frequencies which are of a higher-order of a certain fundamental missing frequency, then an optimal noise can induce a resonance at the missing fundamental frequency. This resonance phenomenon has been called ghost-stochastic resonance [Chialvo {\emph{et al.}}, 2002; Chialvo, 2003; Balenzuela {\emph{et al.}}, 2012]. Ghost-stochastic resonances have been studied in a semiconductor laser [Buld\'{u} {\emph{et al.}}, 2003], two-coupled lasers [Buld\'{u} {\emph{et al.}}, 2005], a monostable Schmidt trigger electronic circuit [Calvo \& Chialvo, 2006], an excitable Chua's circuit [Lopera {\emph{et al.}}, 2006]  and a chaotic Chua's circuit [Gomes {\emph{et al.}}, 2012].

This paper reports our recent investigation on the role of the number of breakpoints $N$ in the modified Chua's circuit model equation on vibrational and ghost resonances. To observe vibrational resonance the circuit is driven by a biharmonic force $f\cos\omega t + g\cos\Omega t$ with $\Omega\gg\omega$ and $\vert f \vert \ll1$. $f\cos\omega t$ is a low-frequency force while $g\cos\Omega t$ is a high-frequency force. We numerically compute the amplitude $A$ of the response of the system at the low-frequency $\omega$ and denote $Q=A/f$ as the response amplitude of the system at the frequency $\omega$. When the control parameter $g$ is varied, $Q$ exhibits a resonance. We analyse the role of the number of breakpoints $N$ on the features of the high-frequency force induced resonance. The number of resonance peaks is found to be $N$. The value of $g$ at which the $i$th resonance occurs and the corresponding value of $Q$ are independent of the number of breakpoints. Further, for each fixed value of $N$, after the last resonance $Q$ does not decay to zero but approaches a nonzero constant value and $Q(g \to \infty)$ scales linearly with $N$. The resonance curve displays hysteresis and a jump phenomenon. That is, the system shows different response curves and sudden jumps in the value of $Q$ when the control parameter $g$ is varied in the forward and reverse directions. We explain these observations using a phase portrait and a basin of attraction plot.

In the vibrational resonance case, the input periodic signal contains only two frequencies $\omega$ and $\Omega$ with $\Omega\gg\omega$. That is, the low-frequency component has only one frequency. We consider the system with multi-frequencies apart from the high-frequency force $g\cos\Omega t$. Specifically, we have chosen the multi-frequency force as 
\begin{equation}
\label{eq1}
    S(t) = f \sum_{i=1}^{n_{\mathrm{f}}}\cos\omega_it,
             \quad \omega_i = (k+i-1) \omega_0,
\end{equation}
with $k \ge 2$ and $\Omega\gg\omega_{n_\mathrm{f}}=(k + n_{\mathrm{f}}-1)\omega_0$. When the system is driven by $S(t)$ in the absence of the high-frequency force ($g=0$), then $Q(\omega_i)\ne0$, $i=1,2,\cdots,n_{\mathrm{f}}$, while $Q(\omega_0)=0$. Resonance occurs at the missing fundamental frequency $\omega_0$, when the amplitude of the high-frequency force is varied. We call this resonance induced by the high-frequency force as \textit{ghost-vibrational resonance}. In this resonance phenomenon there are also $N$ resonance peaks, where $N$ is the number of breakpoints in the sawtooth function of the characteristic curve of the Chua's diode. Furthermore, the jump phenomenon takes place near resonance and we describe it through a phase portrait and a basin of attraction plot. The response amplitude at resonances decreases for increasing values of $k$  and with the number of low-frequency forces.

The outline of the paper is as follows. In Sec.~\ref{vrmcce}, we analyse the features of vibrational resonance in the modified Chua's circuit model equation. We bring out the influence of the number of breakpoints on resonance and account the occurrence of a jump phenomenon near resonance. In Sec.~\ref{rmff}, we study the response of the system in the presence of the multi-low-frequency forces and the high-frequency periodic force. Finally, in Sec.~\ref{concl} we present our conclusions.  

\section{Vibrational Resonance in the Modified Chua's Circuit Equation}
 \label{vrmcce}
\begin{figure}[t]
\begin{center}
\epsfig{figure=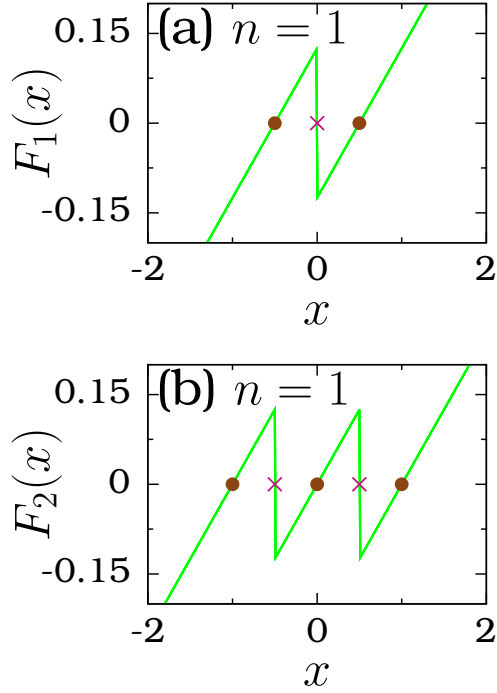, width=0.43\columnwidth}
\end{center}
\caption{Sawtooth functions $F_1(x)$ and $F_2(x)$ given by Eqs.~(\ref{eq2}d) and (\ref{eq2}e) respectively with $n=1,~\xi=0.25$ and $A=0.5$. The solid circles and the symbol `x' mark the locations of the equilibrium points and the breakpoints respectively.}
\label{fig1}
\end{figure}

It is important to analyse the resonance dynamics by varying the number of equilibrium points. There is a simple system in which the number of equilibrium points can be easily varied. The system of our interest with this characteristic property is the modified Chua's circuit model equation [Yu {\emph{et al.}}, 2007], which was introduced to generate multi-scroll attractors. The model equation of this circuit driven by the biharmonic force is
\begin{subequations}
\label{eq2}
\begin{eqnarray}
  \dot{x} & = & \alpha y-\alpha F(x)+S(t)+g\cos\Omega t,\\
  \dot{y} & = & x-y+z,\\ 
  \dot{z} & = & -\beta y,
\end{eqnarray}
where $S(t)=f\cos\omega t$,
\begin{eqnarray}
   F(x) = F_1(x) = \xi x + \xi A{\mathrm{sgn}}(x)
            -\xi A\sum_{j=0}^{n-1}[{\mathrm{sgn}}(x+2jA)
               + {\mathrm{sgn}}(x-2jA)]
\end{eqnarray}
or
\begin{eqnarray}
    F(x) = F_2(x) = \xi x-\xi A\sum_{j=0}^{n-1}
            [{\mathrm{sgn}}(x+(2j+1)A)
              +{\mathrm{sgn}}(x-(2j+1)A)]
\end{eqnarray}
with
\begin{eqnarray}
   {\mathrm{sgn}}(x) =
    \left\{ \begin{array}{rl}
        1, & \;{\mathrm{if}}~x>0 \\          
	0, & \;{\mathrm{if}}~x=0 \\          
	-1, & \;{\mathrm{if}}~x<0 ,
         \end{array}
        \right.
\end{eqnarray}
\end{subequations}
and $\alpha,\;\beta,\;\xi,\;A>0,\;n\ge1$. $F(x)$ given by Eq.~(\ref{eq2}d) or (\ref{eq2}e) is a sawtooth function with amplitude $2A\xi$ and period $2A$. Figures {\ref{fig1}}(a) and {\ref{fig1}}(b) depict the forms of $F_1(x)$ and $F_2(x)$ respectively for $n=1$, $\xi=0.25$ and $A=0.5$. The stable equilibrium points about which scroll orbits occur are given by
%
\begin{eqnarray}
\label{eq3}
    X^* & = & (x^*,y^*,z^*) \nonumber \\
    & = & \left\{ \begin{array}{ll}
          \left[ \pm(2m-1)A,~0,~\mp(2m-1)A \right],
                & \;{\mathrm{for}} 
              ~F_1(x) \\
          \left[0,0,0 \right],~ \left[\pm 2mA,~0,~\mp 2mA
           \right], & \;{\mathrm{for}} ~F_2(x),
         \end{array} \right.
\end{eqnarray}
where $m=1,2,\cdots,n$. The system (\ref{eq2}) with $F_1(x)$ and in the absence of external forces possesses $2n$ (even) number of stable equilibrium points and $N=2n-1$ (odd) number of breakpoints while with $F_2(x)$ it admits $2n+1$ (odd) number of stable equilibrium points and $N=2n$ (even) number of breakpoints. Between two consecutive stable equilibrium points the sawtooth function $F(x)$ has a breakpoint. The number of breakpoints depends on the value of $n$ in Eqs.~(\ref{eq2}d) and (\ref{eq2}e). The breakpoints are given by
\begin{equation}
\label{eq4}
    x_{\mathrm{bp}}^*
       = \left\{ \begin{array}{ll}
	    \pm 2mA, & \;{\mathrm{for}}~F_1(x) \\
            \pm (2m+1)A, & \;{\mathrm{for}}~F_2(x)
         \end{array} \right.
\end{equation}
where $m=0,1,2,\dots,n-1$. Let us analyse the role of the number of breakpoints, $N$, on vibrational resonance.

\subsection{Role of the number of breakpoints $N$ on the resonance}
\begin{figure}[t]
\begin{center}
\epsfig{figure=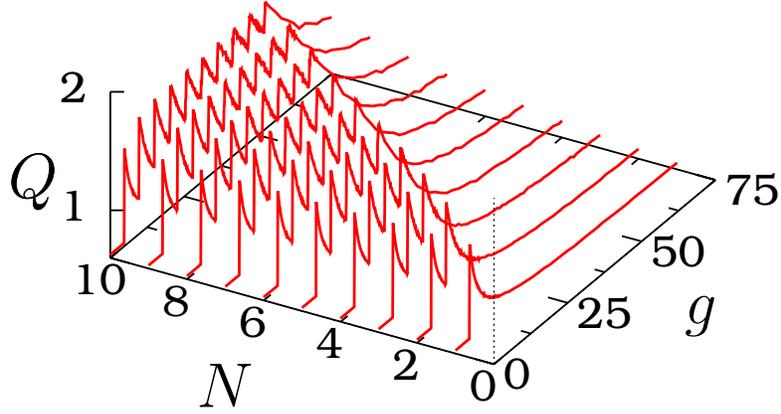, width=0.63\columnwidth}
\end{center}
\caption{Variation of the response amplitude $Q$ as a function of $g$ for various fixed values of the number of breakpoints $N$ of the system (\ref{eq2}). $F(x)$ in Eq.~(\ref{eq2}a) is $F_1$ for odd values of $N$ and $F_2$ for even values of $N$. For each fixed value of $N$, $g$ is varied from $0$ to $75$.}
\label{fig2}
\end{figure}

\begin{figure}[t] 
\begin{center}
\epsfig{figure=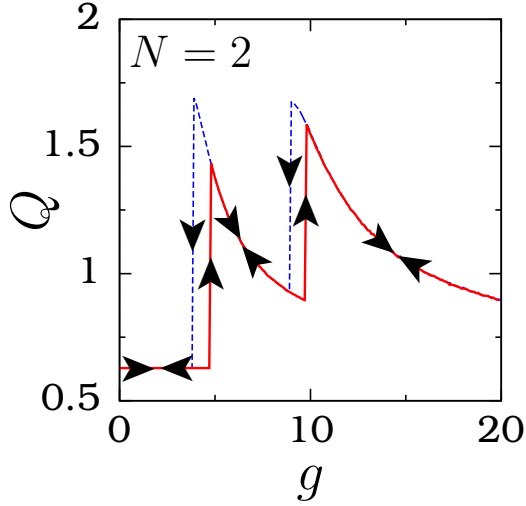, width=0.42\columnwidth}
\end{center}
\caption{Resonance curves obtained by varying $g$ from $0$ to $20$ (continuous curve) and from $20$ to $0$ (dashed curve). We have considered $N=2$ with $F(x)$ in Eq.~(\ref{eq2}a) as $F_2(x)$ given by Eq.~(\ref{eq2}e). The arrows indicate the path followed by the response amplitude when $g$ is varied in the forward and reverse directions.}
\label{fig3}
\end{figure}

The determination of an approximate theoretical expression for the response amplitude in Eq.~(\ref{eq2}) is not an easy task, so that we have decided to perform a numerical simulation. From the numerical solution of $x(t)$, the response amplitude $Q$ is computed through $Q=\sqrt{Q_{\mathrm{s}}^2+Q_{\mathrm{c}}^2}/f$ where
\begin{subequations}
 \label{eq5}
  \begin{eqnarray}
   Q_{\mathrm{s}} 
     & = & \frac{2}{MT}\int_0^{MT}x(t)\sin\omega t~dt,\\
   Q_{\mathrm{c}}
     & = & \frac{2}{MT}\int_0^{MT}x(t)\cos\omega t~dt,
  \end{eqnarray}
\end{subequations}
where $T={2\pi}/{\omega}$ and $M$ is taken as $500$. Here, we have fixed the values of the parameters in Eq.~(\ref{eq2}) as $\alpha=6$, $\beta=16$, $A=0.5$, $\xi=0.25$, $f=0.1$, $\omega=1$ and $\Omega=10\omega$. Figure \ref{fig2} presents the numerically computed response amplitude $Q$ versus $g$ for several fixed values of $N$.
%
%
Vibrational resonance occurs when the control parameter $g$ is varied. In obtaining the Fig.~\ref{fig2}, for each fixed value of $N$, $g$ is varied from $0$ to $75$ in steps of $0.01$. The system of Eq.~(\ref{eq2}) is numerically integrated using a fourth-order Runge--Kutta method with step size $(2\pi/\omega)/1000$. The first $10^3$ drive cycles are left as transient and the values of $x(t)$ corresponding to the next $500$ drive cycles are used to compute the response amplitude. For $g=0$, the initial condition is chosen in the neighbourhood of the origin. For other values of $g$, the initial condition is taken as the last value of $(x,y,z)$ of the previous value of $g$. When $g$ is decreased from a large value, say $75$, $Q$ is found to follow a different path near each resonance. An example is shown in Fig.~\ref{fig3} for $N=2$.  In this figure, solid and dashed curves represent the resonance curve obtained when $g$ is varied in the forward and reverse directions respectively.
%
%
Before explaining the resonance curves in Fig.~\ref{fig3}, we point out the influence of the number of breakpoints $N$ on the resonances.
%
\begin{figure}[!h]
\begin{center}
\epsfig{figure=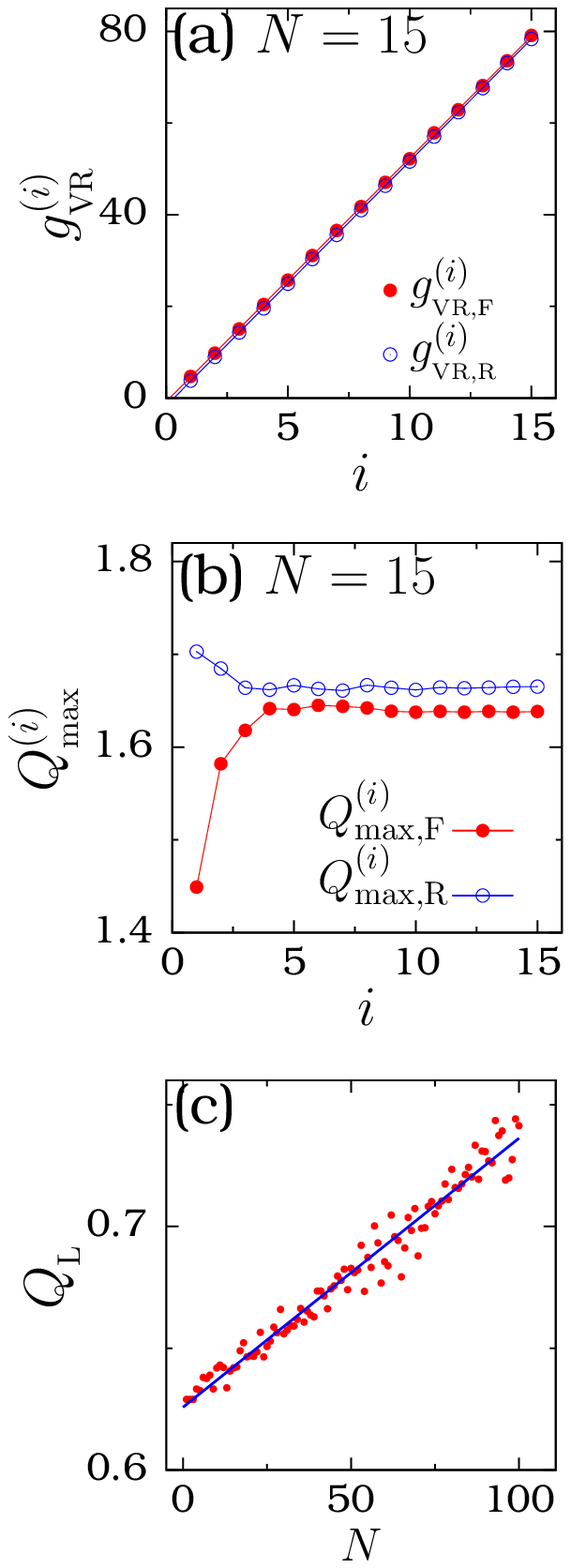, width=0.42\columnwidth}
\end{center}
\caption{Plots of (a) $g^{(i)}_{_\mathrm{VR}}$, the value of $g$ at which $Q$ becomes maximum, versus $i$ (the number of resonance) and (b) $Q_{\mathrm{max}}^{(i)}$, the value of $Q$ at $g=g^{(i)}_{_\mathrm{VR}}$, versus $i$ for $N=15$. (c) Variation of the limiting value of $Q$ (the value of $Q$ at $g=2000$) with the number of breakpoints $N$. In the subplots (a) and (c) the continuous lines are the best straight-line fits to the numerical data.}
\label{fig4}
\end{figure}
%
%
\begin{itemize}
 \item 
For each fixed value of $N$ there are $N$ resonances.
 \item 
In both cases of varying $g$ in the forward direction from $0$ and in the reverse direction from a large value, for convenience, we number the resonance peaks from left-side to right-side as $1,2,\cdots,N$. In the former case, we denote the values of $g$ at which the $i$th resonance occurs as $g_{_\mathrm{VR,F}}^{(i)}$ and the corresponding value of the response amplitude as $Q_{\mathrm{max,F}}^{(i)}$, while in the latter case these two values are denoted as $g_{_\mathrm{VR,R}}^{(i)}$ and $Q_{\mathrm{max,R}}^{(i)}$ respectively. The values of $g_{_\mathrm{VR}}^{(i)}$ and $Q_{\mathrm{max}}^{(i)}$ are independent of $N$. This is because the shapes of $F(x)$ between various sets of successive equilibrium points are the same.
 \item 
The values of $g_{_\mathrm{VR}}^{(i)},~i=1,2,\cdots,N$ are equally spaced for each fixed value of $N$. Furthermore, $g_{_\mathrm{VR}}^{(i)}$ varies linearly with $i$ as shown in Fig.~\ref{fig4}(a) following the relation $g_{_\mathrm{VR,F}}^{(i)}=5.304i-0.615$ and $g_{_\mathrm{VR,R}}^{(i)}=5.334i-1.686$.
\item 
For each value of $N$, $Q_{\mathrm{max,F}}^{(i)}$, $i=1,2,\cdots,N$ increases with $i$ while $Q_{\mathrm{max,R}}^{(i)}$ decreases with $i$. However, both of them approach almost the same constant value. This is shown in Fig.~\ref{fig4}(b) for $N=15$.
 \item 
In a typical monostable and bistable polynomial potential ($V(x)$) systems, where $V(x)\to\infty$ as $\vert x\vert\to\infty$, the response amplitude decays to zero with the control parameter $g$ far after the last resonance  [Landa \& McClintock, 2000; Jeyakumari {\emph{et al.}}, 2009]. In contrast to this fact, $Q$ approaches a nonzero limiting value for the  system of the Eq.~(\ref{eq2}). In this connection we note that $F_1(x)$ and $F_2(x)$ are different from the above mentioned polynomial potential. $F_1$ and $F_2\to\pm\infty$ linearly as $x\to\pm\infty$. Further, $F_1$ and $F_2$ are piece-wise linear between the two end breakpoints. The mechanism of vibrational resonance in these two types of systems are different. We denote $Q_{_\mathrm{L}}$ as the value of $Q$ computed at a sufficiently large value of $g$, say, $2000$. Figure \ref{fig4}(c) shows the variation of $Q_{\mathrm{L}}$ with $N$. Furthermore, $Q_{\mathrm{L}}$ varies linearly with $N$ as $Q_{\mathrm{L}}=0.62581+0.001N$.
\item
For a fixed $N$, when the distance $D$ between the first and last breakpoints is varied then $Q_{\mathrm{L}}$ (calculated at $g=2000$) is found vary. $Q_{\mathrm{L}}$ can be varied by varying $N$ for fixed $D$ or varying $D$ by fixing $N$.
\item
For a fixed value of $N$, for $g$ value less than a critical value the orbit lies within the left-most and right-most breakpoints. This critical value of $g$ depends on $N$. Suppose consider, a range of $N$ values in the interval $N'$ and $N''$ with $N''>N'$. For a value of $g$ if the orbit lies within the left-most and right-most breakpoints corresponding to $N=N'$ then the orbit remains the same for all values of $N$ in the interval $[N',N'']$. The width of the orbit $x_{\mathrm{w}}$ defined as $x_{\mathrm{max}}-x_{\mathrm{min}}$ remains the same. For sufficiently large value of $g$ for which the orbit for various values of $N$ are all visit the regions outside the left- and right-most breakpoints, the quantity $x_{\mathrm{w}}$ is found to increase linearly with $N$. This is a reason for linear variation of $Q_{\mathrm{L}}$. That is, the size of the orbit is influenced by the end breakpoints. If an analytical expression for $Q$ is known then we can able to explicitly identify the way in which $N$ or the end breakpoints influence the value of $Q$ for very large $g$. But for the system (\ref{eq2}) theoretical calculation of $Q$ is very difficult.
\end{itemize}
\subsection{Jump phenomenon}
\begin{figure}[t]
\begin{center}
\epsfig{figure=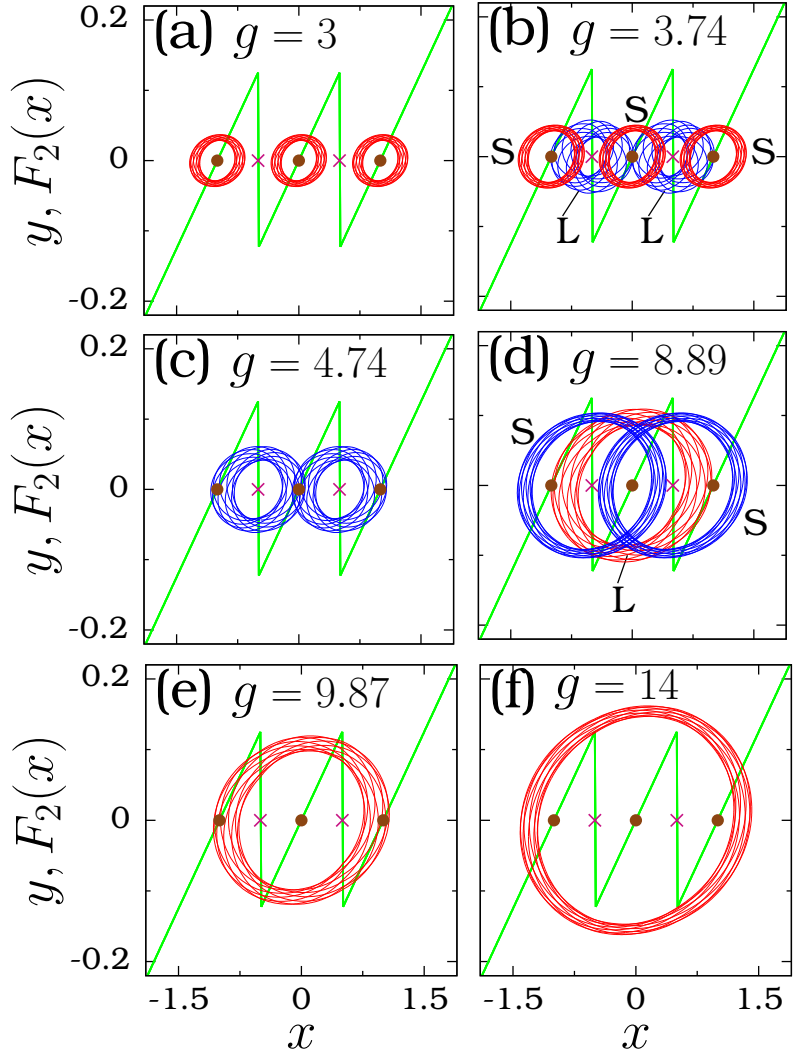, width=0.7\columnwidth}
\end{center}
\caption{Phase portraits of the coexisting orbits of the system Eq.~(\ref{eq2}) with $N=2$ in the $x-y$ plane for various fixed values of $g$. The solid circles and `x' mark the equilibrium points $x^*=0$, $\pm 1$ and the breakpoints $x_{\mathrm{bp}}^*=\pm 0.5$ respectively. $F_2(x)$ is also shown for illustrating the influence of the high-frequency periodic force. In the subplots (b) and (d), the value of $Q$ for the orbits S is smaller than that of the orbits labelled as L.}
\label{fig5}
\end{figure}

We explain the resonance curves (solid and dashed curves) in Fig.~\ref{fig3} corresponding to $N=2$. In this case, the function $F(x)$ in Eq.~(\ref{eq2}) is $F_2(x)$ given by Eq.~(\ref{eq2}e). In the absence of the biharmonic force, the system has three stable equilibrium points $(x^*=0,\pm1)$ and two breakpoints $x_{\mathrm{bp}}^*=\pm0.5$. When the amplitude of the high-frequency force is varied, we can clearly notice the occurrence of hysteresis and jumps in the value of the response amplitude in Fig.~\ref{fig3}. In order to get more insight on the resonance curve, we present the phase portrait of orbits in the $x-y$ plane along with $F_2(x)$ for six values of $g$ in Fig.~\ref{fig5}. For $0<g<3.74$ there are three coexisting periodic orbits of period-$T(=2\pi/\omega)$, one about each of the stable equilibrium points $x^*=0,\pm1$ as shown in Fig.~\ref{fig5}(a) for $g=3$. These orbits do not cross the barriers at the breakpoints. The $Q$ of these orbits are all the same and $\ll1$. There are no stable orbits about the breakpoints $x_{\mathrm{bp}}^*=\pm0.5$.

At $g=3.74$, two more stable periodic orbits of period-$T$ centered about the two breakpoints $x_{\mathrm{bp}}^*$ are born, in addition to the three orbits centered about $x^*$. The five coexisting orbits are displayed in Fig.~\ref{fig5}(b). All these orbits coexist for $3.74 \le  g < 4.74$. The $Q$ of the newly born two orbits (marked as L in Fig.~\ref{fig5}(b)) are equal and much higher than that of the other three orbits. For example, at $g=3.74$, the $Q$ of the orbits centered about the equilibrium points (marked as S in Fig.~\ref{fig5}(b)) and the breakpoints (marked as L in Fig.~\ref{fig5}(b)) are $0.628$ and $1.693$ respectively. At $g=4.74$, the S orbits centered about $x^*$ disappear and only the two orbits centered about $x_{\mathrm{bp}}^*$ coexist (see Fig.~\ref{fig5}(c)). Consequently, when $g$ is varied in the forward direction the response amplitude jumps from a lower value to a higher value at $g=4.74$.

As $g$ increases further from the value $4.74$, the orbits expand and $Q$ at the frequency $\omega$ decreases. For $4.74\le g<8.89$, the two orbits centered about $x_{\mathrm{bp}}^*$ coexist alone and moreover they enclose two equilibrium points one to the left-side and the other one to the right-side of the breakpoint. At $g=8.89$ the high-frequency force gives birth to a periodic orbit centered about the equilibrium point $x^*=0$ and enclosing all the breakpoints. We note that at $g=3.74$ the newly born two orbits are centered about breakpoints. At $g=8.89$ the response amplitude of the newly created orbit is $1.67$, while that of the other two coexisting orbits is $0.92$. The three orbits coexist for $8.89 \le g < 9.87$. In Fig.~\ref{fig3} for $g \in [8.89,9.87]$ the response amplitude curves represented by continuous and dashed lines correspond to the orbits centered about the breakpoints and the newly born orbit centered about the equilibrium point $x^*=0$ respectively.  In obtaining Fig.~\ref{fig3} $g$ is increased with step size $0.01$ in the forward direction from $0$ with $(x(0),y(0),z(0))$ for $g=0$ chosen  in the neighbourhood of the origin while for other values of $g$ the value of  $(x(0),y(0),z(0))$ is the last value of $(x,y,z)$ of the previous value of $g$. As a result in the numerical simuation of obtaining  Fig.~\ref{fig3}, in the interval $8.89 \le g < 9.87$ the orbits centered about the breakpoints are realized since these orbits alone exist just below $g=8.89$ and $(x(0),y(0),z(0))$ is on the basin of attraction of these orbits. Consequently, in Fig.~\ref{fig3} the resonance curve traces the lower branch (continuous curve) corresponding to the orbits centered about the breakpoints.  (The upper branch curve (dashed curve) is obtained when $g$ is varied in the reverse direction from the value, say, $20$.).  The small amplitude orbits centered about the breakpoints become unstable at $g=9.87$.  Therefore, we observe a jump in $Q$ from a lower value to a higher value in Fig.~\ref{fig3}. The value of $Q$ decreases with further increase in $g$ and there is no more birth of a periodic orbit as shown in Figs.~\ref{fig5}(e) and (f).

From the above, we can generalize the effect of the high-frequency force in the system with $l$ breakpoints and $l+1$ equilibrium points. The resonance curve would display $l$ jumps. We denote the values of $g$ at which jumps in $Q$ occur when $g$ is varied in the forward direction as $g_{\mathrm{j}i}$, $i=1,2,\cdots,l$ with $ g_{\mathrm{j1}} < g_{\mathrm{j2}} <$ $\cdots$ $< g_{\mathrm{j}l}$.  We noticed that in certain intervals of $g$ orbits centered about the equilibrium points and orbits centered about the breakpoints coexist.  We denote the starting value of $g$ of $i$th  such an interval as $g_{\mathrm{b}i}$.    We note that  $g_{\mathrm{b}i} < g_{\mathrm{j}i}$. For $g<g_{\mathrm{b}1}$ there are $l+1$ orbits centered about the equilibrium points. At $g=g_{\mathrm{b}i}$, $i=\,$odd (even) $l+1-i$ orbits born with each one centered about the $l+1-i$ breakpoints (equilibrium points) with the value of $Q$ higher than those of the $l+2-i$ coexisting orbits centered about the equilibrium points (breakpoints). These two set of orbits coexist for $g_{\mathrm{b}i} < g < g_{\mathrm{j}i}$, $i=\,$odd (even).  At $g_{\mathrm{j}i}$, $i=\,$odd (even) the orbits centered about the equilibrium points (breakpoints) become unstable. For $g_{\mathrm{j}i} \le g < g_{\mathrm{b}(i+1)}$, $i=\,$odd (even), in the numerical simulation we realize only the orbits centered about the breakpoints (equilibrium points). When $g$ is varied in the forward direction from a small value, jumps in the response amplitude $Q$ occur at $g_{\mathrm{j}i}$ from a lower value to a higher value. If $g$ is decreased from a value of $g \gg g_{\mathrm{j}l}$ then jumps in $Q$ occur at $g_{\mathrm{b}i}$ from a higher value to a lower value.
\subsection{Basin of attraction for $3.74 \le g \le 4.74$}
\begin{figure}[t]
\begin{center}
\epsfig{figure=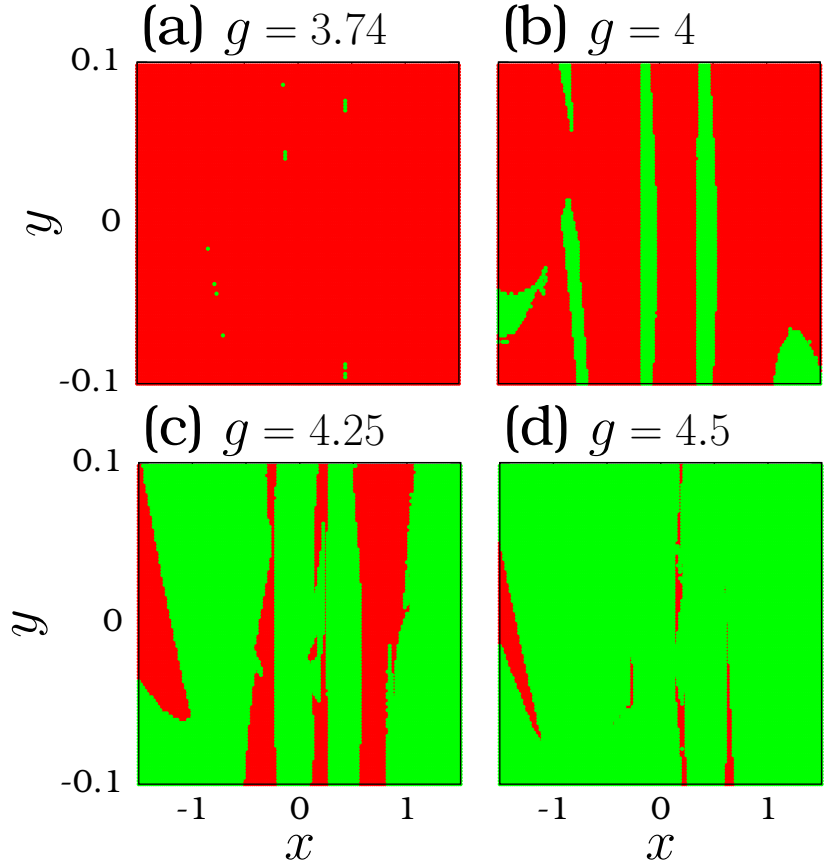, width=0.7\columnwidth}
\end{center}
\caption{Change in the basin of attraction of large response amplitude orbits (marked as L in Fig.~\ref{fig5}(b)) (green colored region) and small amplitude orbits (marked as S in Fig.~\ref{fig5}(b)) (red colored region) as a function of the parameter $g$ in $x-y$ plane with $z=-0.5$.}
\label{fig6}
\end{figure}

In Fig.~\ref{fig5} we found that when $g$ is varied from a small value, then for $0 < g <3.74$ there are only three small amplitude orbits about the equilibrium points $x^*$, while for $3.74 \le g < 4.74$ in addition to these orbits there are two more stable orbits centered about the breakpoints $x_{\mathrm{bp}}^*$. The three orbits centered about $x^*$ disappear at $g = 4.74$. In order to further check the disappearance of these three orbits, we numerically calculate the basin of attraction of the orbits centered about the equilibrium points and the breakpoints.  We consider the region $x \in [-1.5,1.5]$, $y \in [-0.1,0.1]$ with $z=-0.5$ and divide this region into $150 \times 150$ grid points. With each grid point as an initial condition, we numerically integrate Eq.~(\ref{eq2}) and after a sufficient transient, we identify whether the trajectory is on any  one of the three orbits centered about $x^*$ (small amplitude orbits) or on any one of the two orbits centered about $x_{\mathrm{bp}}^*$ (large amplitude orbits). In the $x-y$ plane, we mark the initial conditions which approach any one of the large amplitude orbits by green color and red color for the initial conditions approaching small amplitude orbits. Figure \ref{fig6} presents the plot of the basin of attraction of small and large amplitude orbits for four values of $g$. 
For $g=3.74$, at which the large amplitude orbits are born, the area of the basin of attraction of the large amplitude orbits is very small, while that of the small amplitude orbits is very large. The area of the basin of attraction of the large (small) amplitude orbits increases (decreases) with $g$. This is clearly seen in Figs.~\ref{fig6}(b-d). We define $N_{\mathrm{S}}~(N_{\mathrm{L}})$ as the ratio of the number of initial conditions that approach the small (large) amplitude orbits and the total number of initial conditions. We compute $N_{\mathrm{S}}$ and $N_{\mathrm{L}}$ for $3.74<g<4.74$. As $g$ increases from $3.74$ to $4.74$, the quantity $N_{\mathrm{S}}$ decays to zero while $N_{\mathrm{L}}$ increases from a small value and becomes $1$ at $4.74$.

\section{Resonance with a Multi-Frequency Force}
 \label{rmff}
In this section, we consider the system (\ref{eq2}) with $S(t)$ in Eq.~(\ref{eq2}a) as the multi-frequency input signal given by Eq.~(\ref{eq1}). We choose the values of the parameters in Eq.~(\ref{eq2}) as $\alpha=4$, $\beta=14$, $A=0.5$, $\xi=0.25$, $\omega_0=0.1$, $f=0.04$, $\Omega=20\omega_0$. When $k=2$, $n_{\mathrm{f}}=2$ and $g=0$ the frequencies present in the input signal are $\omega_1=2\omega_0$ and $\omega_2=3\omega_0$. The frequency $\omega_0$ is absent. The output signal $x(t)$ of the system (\ref{eq2}) for each fixed value of $N$ contains components with the frequencies $\omega_1=2\omega_0$ and $\omega_2=3\omega_0$. The frequency $\omega_0$ is not found in the output signal. The interval of $g$ where $\omega_0$ is absent generally depends on the values of the other parameters of the system and the parameters in the input signal. We show the occurrence of a resonance at the missing frequency $\omega_0$ when $g$ is varied and explore the influence of the number of breakpoints $N$ on it.

\subsection{Ghost-vibrational resonance}
%
\begin{figure}[t]
\begin{center}
\epsfig{figure=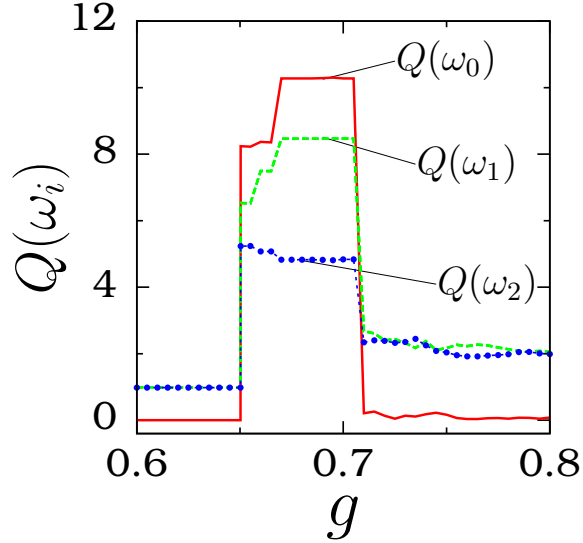, width=0.5\columnwidth}
\end{center}
\caption{Numerically computed response amplitudes $Q$ at the frequencies $\omega_0$, $\omega_1$ and $\omega_2$ as a function of the amplitude $g$ of the high-frequency force. The system (\ref{eq2}) is subjected to the external force $S(t)$ given by Eq.~(\ref{eq1}) with $n_{\mathrm{f}}=2$, $k=2$, $\omega_0=0.1$ and $f=0.04$. The values of the other parameters in Eq.~(\ref{eq2}) are $\alpha=4$, $\beta=14$, $A=0.5$, $\xi=0.25$ and $\Omega=20\omega_0$. The number of breakpoints $N$ in $F(x)$ is $1$, that is, $F(x)=F_1(x)$ (Eq.~(\ref{eq2}d)) with $n=1$.}
\label{fig7}
\end{figure}

\begin{figure}[t]
\begin{center}
\epsfig{figure=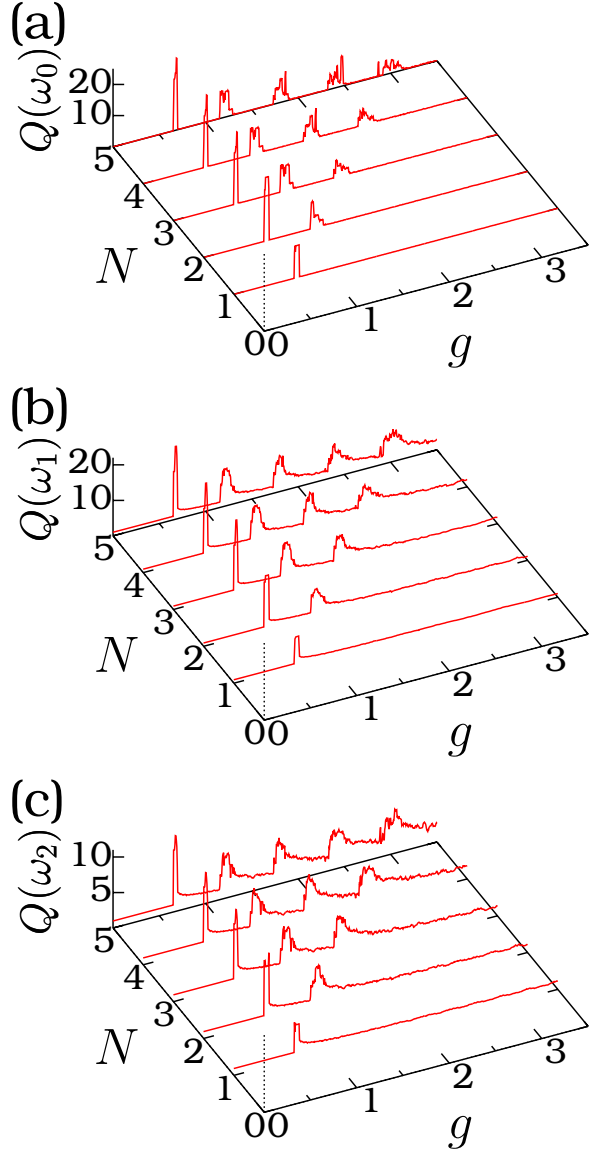, width=0.48\columnwidth}
\end{center}
\caption{Variation of (a) $Q(\omega_0)$, (b) $Q(\omega_1)$ and (c) $Q(\omega_2)$ versus $g$ for the number of breakpoints $N=1,2,\cdots,5$ for the system (\ref{eq2}) with $S(t)$ given by Eq.~(\ref{eq1}). Here $n_{\mathrm{f}}=2$, $k=2$ and $\omega_0=0.1$.}
\label{fig8}
\end{figure}
%
Figure \ref{fig7} depicts the variation of the response amplitudes $Q(\omega_0)$, $Q(\omega_1)$ and $Q(\omega_2)$ with the control parameter $g$ (varied from zero) when the number of breakpoints is $N=1$. For $g < 0.6502$ and $g > 0.709$, the response amplitude $Q(\omega_0) \approx 0$ while $Q(\omega_1)$ and $Q(\omega_2)$ are nonzero. Near $g=0.6502$ and $0.709$ a sharp variation of $Q$ at these three frequencies $\omega_0$, $\omega_1$ and $\omega_2$ takes place. For $g \in [0.6502,0.709]$, $Q(\omega_0)$ is greater than $Q(\omega_1)$ and $Q(\omega_2)$. The point is that the output signal of the system has a periodic component with the frequency $\omega_0$, which is missing in the input signal. The missing frequency is termed as ghost frequency [Chialvo {\emph{et al.}}, 2002; Chialvo, 2003]. In Fig.~\ref{fig7} we observe that not only $Q(\omega_0)$ exhibits a resonance, but actually is the dominant resonance. As this resonance at the missing frequency (in the input signal) is induced by the high-frequency force, we call it ghost-vibrational resonance. The resonance displayed by the response amplitudes $Q(\omega_1)$ and $Q(\omega_2)$ are the well known vibrational resonance.

When the size of the stable orbit smoothly varies with the control parameter $g$ then $Q$ would vary smoothly and in this case $Q$ would display a peak at resonance. Differently, in Fig.~\ref{fig7} we observe abrupt jumps in $Q$ when $g$ is increased. This indicates that the size of stable orbit is robust with the variation of $g$. This type of behaviour is found in excitable systems for certain parametric values.

In Fig.~\ref{fig8}, we plot $Q(\omega_0)$, $Q(\omega_1)$ and $Q(\omega_2)$ versus $g$ for $N=1,2,\cdots,5$ with the starting value of $g$ as $0$. Compare Fig.~\ref{fig8} with Fig.~\ref{fig2} (corresponding to the single low-frequency force, $n_{\mathrm{f}}=1,$ in $S(t)$ given by Eq.~(\ref{eq1}). There are some similarities and differences in the variation of the response amplitudes. In Fig.~\ref{fig8}, the number of resonances is equal to the number of breakpoints $N$ as in Fig.~\ref{fig2}. In Fig.~\ref{fig2}, $Q_{\mathrm{L}}(\omega)$, the value of $Q$ in the limit of $g \to \infty$, is nonzero while in Fig.~\ref{fig8}, $Q(\omega_0, g \to \infty) \approx 0$. $Q_{\mathrm{L}}(\omega_1)$ and $Q_{\mathrm{L}}(\omega_2)$ does not decay to zero in the limit of $g \to \infty$. $Q(\omega_0) \approx 0$ for a range of values of $g$ between two successive resonances. For large values of $N$, as the number of resonance increases, the value of $Q(\omega_0)$ at resonance decreases, while in Fig.~\ref{fig2} $Q$ at successive resonance does not decay but approaches a nonzero constant value.

\subsection{Hysteresis and jump phenomenon}
\begin{figure}[t]
\begin{center}
\epsfig{figure=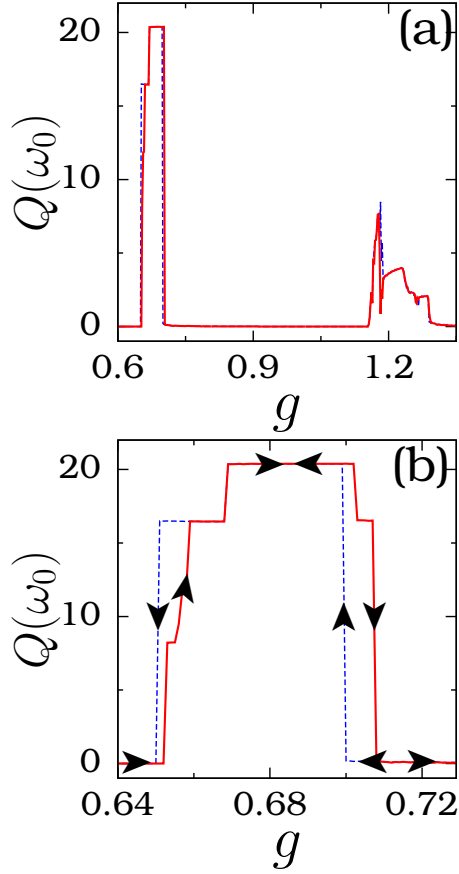, width=0.4\columnwidth}
\end{center}
\caption{(a) $Q(\omega_0)$ versus $g$ for the system (\ref{eq2}) driven by $S(t)$ given by Eq.~(\ref{eq1}) with $n_{\mathrm{f}}=2$, $k=2$ and $\omega_0=0.1$. Here $N=2$. (b) Magnification of $Q(\omega_0)$ around the first resonance. In both subplots the solid and dashed curves represent the resonance curves obtained when $g$ is varied from $0$ to $1.5$ and $1.5$ to $0$ respectively. The path followed by the response amplitude is indicated by the arrows.}
\label{fig9}
\end{figure}

The jump phenomenon and hysteresis found in the case of the system (\ref{eq2}) with $S(t)=f\cos\omega t$ are also observed when $S(t)$ is a multi-frequency force. For example, Fig.~\ref{fig9}(a) shows $Q(\omega_0)$ versus $g$ for $k=2$, $n_{\mathrm{f}}=2$, $N=2$. The solid and dashed curves represent the resonance curve obtained by varying $g$ in forward and reverse directions respectively. Since hysteresis is not clearly visible in Fig.~\ref{fig9}(a), we present the magnification of variation of $Q(\omega_0)$ around the first resonance region in Fig.~\ref{fig9}(b). $Q(\omega_0)$ assumes two different values in the intervals $g \in [0.6502,0.66]$ and $g \in [0.7,0.709]$. Hysteresis is observed with $Q(\omega_1)$ and $Q(\omega_2)$ also.

\begin{figure}[!h]
\begin{center}
\epsfig{figure=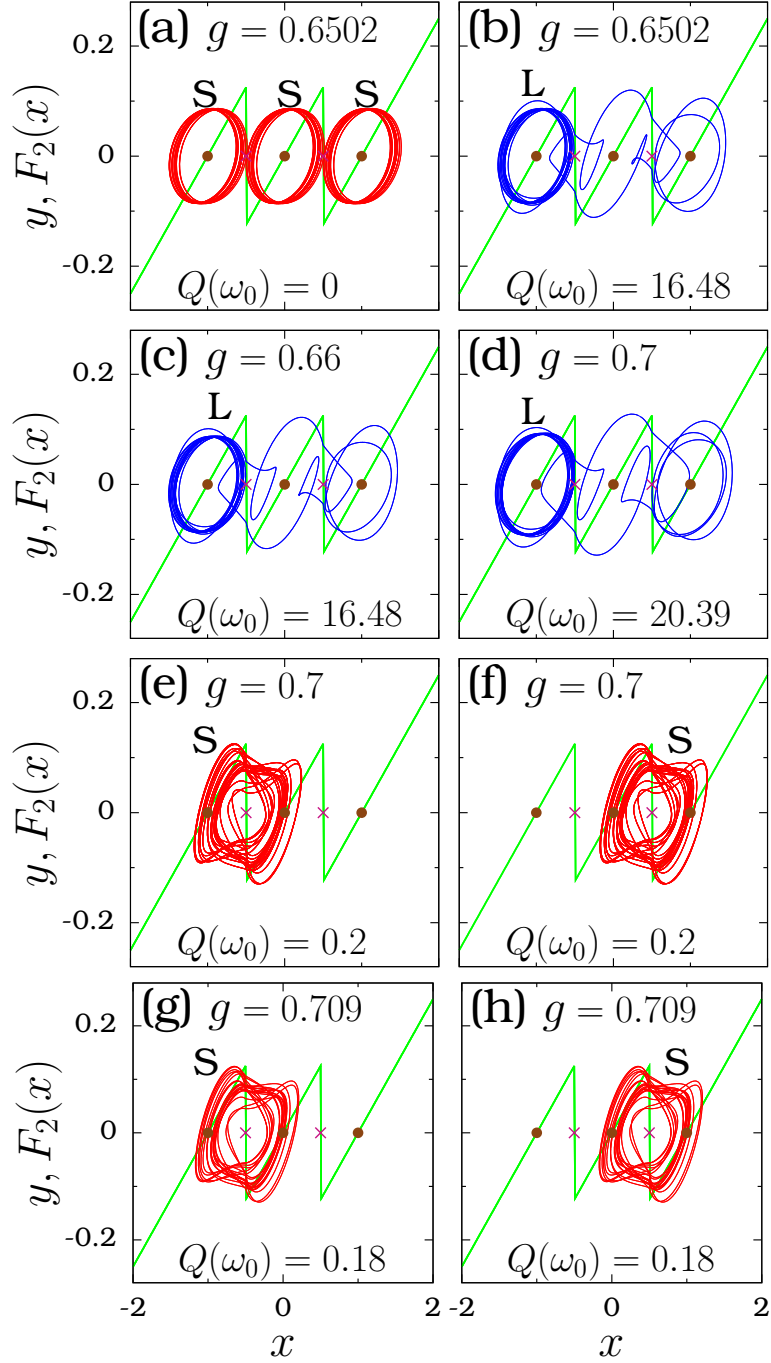, width=0.73\columnwidth}
\end{center}
\caption{Phase portraits of the orbits present at the starting of the first hysteresis (subplots a and b), at the end of the first hysteresis (c), at the starting of  the second hysteresis (d-f) and at the end of the second hysteresis (g and h). The solid circles and the symbol `x' denote the equilibrium points and the breakpoints respectively. $F(x) = F_2(x)$ is also plotted.}
\label{fig10}
\end{figure}

In Fig.~\ref{fig9}(b), $Q(\omega_0)$ is single-valued for the values $0 < g < 0.6502$.  As pointed out in Sec.~II, in the absence of external forces there are three equilibrium points $x^*=0$, $\pm 1$ and there are two breakpoints $x_{\mathrm{bp}}^* = \pm 0.5$ in between two of them. For $g < 0.6502$, three periodic orbits coexist centered about the equilibrium points and moreover for these orbits $Q(\omega_0) = 0$. We denote these three orbits as S. At $g = 0.6502$ in addition to these three orbits another orbit enclosing all the three equilibrium points is born. $Q(\omega_0)$ of this new orbit is much higher than that of the S orbits. We designate this orbit as L. The phase portraits of the three S orbits and one L orbit are shown in Figs.~\ref{fig10}(a) and (b), respectively. 
These two distinct set of orbits coexist for $0.6502 \le g < 0.66$. At $g = 0.66$, the three S orbits disappear and the orbit L alone exists (see Fig.~\ref{fig10}(c)). For $0.66 < g < 0.7$ the response amplitude curve is single-valued and the value of $Q(\omega_0)$ is that of the orbit L. At $g = 0.7$ the S orbits reappear with their centers being the breakpoints. Since the number of breakpoints is two, the number of S orbits is also two. $Q(\omega_0)$ of $S$ orbits is $0.2$ while that of the L orbit is $20.39$. The two S orbits and the L orbit are shown in Figs.~\ref{fig10}(d-f).  The orbit L disappears at $g = 0.709$. The phase portraits of the two S orbits at $g = 0.709$ are shown in Figs.~\ref{fig10}(g) and (h). We call the interval $0.6502 \le g \le 0.709$ the first resonance region.

The second resonance region is $1.163 \le g \le 1.3$. In this interval of $g$, also two hystereses are found. For $0.709 \le  g < 1.163$ two S orbits coexist. The L orbit reappears at $g=1.163$. Consequently, there are three coexisting  orbits for $1.163 \le  g < 1.169$.  Next, at $g = 1.169$ the two S orbits disappear. For $1.169 \le g < 1.179$ the response amplitude traces the $Q(\omega_0)$ of the L orbit. At $g = 1.179$ one S orbit alone is reborn with its center being shifted to the middle equilibrium point enclosing all the three equilibrium points. Then the L orbit disappears at $g = 1.186$ and for $g \ge 1.186$ the S orbit alone is present.

We generalize the above results for $N>1$. There are two hystereses in each of the resonance intervals of $g$. The total number of hystereses is $2N$. (The number of hystereses in the case of a single low-frequency force is always $N$ as shown in Fig.~\ref{fig2}). There are $N+1$ equilibrium points and $N$ breakpoints. We denote the left-most equilibrium point as $x_1^*$, the next one as $x_2^*$, and so on. For each value  of $N$, $N > 1$, before the starting of the first hysteresis, $N+1$ number of S orbits coexist with one about each $x^*$. At the starting of the first hysteresis $N-1$ number of L orbits are born. The first of these encloses $x_1^*$, $x_2^*$ and $x_3^*$, second L orbit encloses $x_2^*$, $x_3^*$ and $x_4^*$ and so on. At the end of the first hysteresis all the S orbits disappear and the L orbits alone are present. At the beginning of the second hysteresis, $N$ number of S orbits reappear with their centers being shifted to the breakpoints. The $N-1$ L orbits disappear at the end of the second hysteresis.  Then at the beginning of third and other higher odd numbered hystereses intervals the number of L orbits born decreases by one while the number of equilibrium points enclosed by them increases by one.  This process continues until the number of L orbits born become one (this orbit encloses all the equilibrium points). All the L orbits disappear at the end of each even numbered  hysteresis.

What happens to the S orbits?  The S orbits disappear at the end of an odd number of hysteresis. At the beginning of a consecutive even number of hysteresis, the number of S orbits reborn decreases by one and further the center is shifted between the equilibrium points and the breakpoints. The number of equilibrium points enclosed by an S orbit increases when the number of hysteresis increases. 

At the beginning of the last hysteresis there are only two coexisting orbits: one L orbit and one S orbit. At the end of the last hysteresis, the L orbit disappears. For $g$ values beyond the last hysteresis, only one S orbit exists. In each resonance interval the large value of $Q$ is associated with the L orbits. 

\begin{figure}[t]
\begin{center}
\epsfig{figure=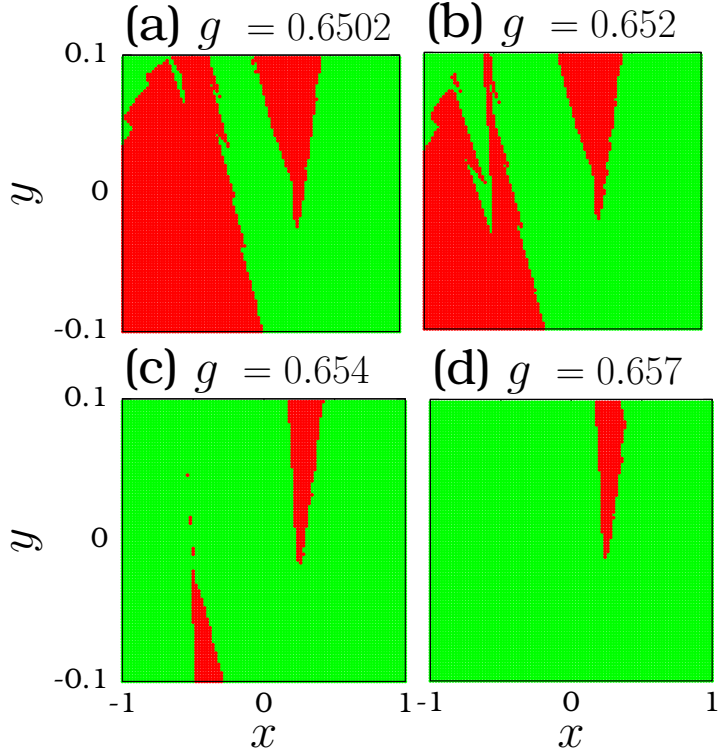, width=0.6\columnwidth}
\end{center}
\caption{Change in the basin of attraction of the L orbits (green colored region) and the S orbits (red colored region) for a few fixed values of $g$ near the end of the first hysteresis in the first resonance region (shown in Fig.~\ref{fig9}(b)).  The region $x \in [-1,1]$ and $y \in [-0.1,0.1]$ with $z=0$ is divided into $150 \times 150$ grid points.}
\label{fig11}
\end{figure}

In Fig.~\ref{fig10}(b) in the first hysteresis interval $0.6502 < g < 0.66$, one L orbit and three S orbits are present for $N=2$. At $g = 0.66$ the three S orbits disappear. In this interval, as $g$ increases from $0.6502$ the volume of the basin of attraction of the S orbits decreases and vanishes at $g = 0.66$.  Figure \ref{fig11} displays the numerically computed basin of attraction of the L orbit (green colored region) and the S orbit (red colored region) for four values of $g$ in the first hysteresis interval. We compute $N_{\mathrm{S}}$ ($N_{\mathrm{L}}$), the ratio of the number of initial conditions approaching the S orbits (the L orbits) for several values of $g$. $N_{\mathrm{S}} \to 0$ while $N_{\mathrm{L}} \to 1$ as $g \to 0.66$ from $0.6502$.

\subsection{Dependence of $Q(\omega_0)$ on $N$ in the resonance intervals}
Another interesting result observed from Fig.~\ref{fig8} is that $Q(\omega_0)$ versus $g$ in the interval $0 < g < g_l$, $l < N$ (where $g_l$ is the end value of $g$ of the $l$th resonance interval) for $N=l+1,l+2,\cdots$ is the same. To account for this result, consider the case $l = 1$. For a fixed value of $g$ in the interval $0 < g < g_1$ ($ = 0.709$), the shapes of all the S orbits and the L orbits present are independent of the value of $N$, $N > 1$. The differences are in the values of $x^*$ ($x$-component of the equilibrium points) about which they occur and the values of $x_{\mathrm{min}}$ and $x_{\mathrm{max}}$ where $x_{\mathrm{min}}$ and $x_{\mathrm{max}}$ are the minimum and maximum values respectively of $x(t)$ of the orbits. However, $x_{\mathrm{max}} - x_{\mathrm{min}}$ remains the same. As a result, $Q(\omega_0)$ versus $g$ ($0 < g < g_1$) for $N = 2, 3, \cdots$ are the same. The cases $l = N$ and $N = 1$ are excluded.

We point out the results for the last resonance interval $(g_{_{N-1}} < g < g_{_{N}})$ for $N > 1$ and the resonance interval for $N = 1$. In the last resonance interval for $N > 1$ an S orbit present encloses $N$ equilibrium points while the L orbit  encloses all the $N+1$ equilibrium points.  That is, the number of equilibrium points enclosed by the S orbit (as well as the L orbit)  increases with increase in the value of $N$.  Hence the phase portraits of the S orbits (and also that of the L orbits) are different for different values of $N$.  Consequently, $Q(\omega_0)$ versus $g$ in the last resonance region for $N > 1$ are not the same.

For all the values of $N$, the first resonance interval is $[0.6502 < g < 0.709]$.  However, $Q(\omega_0)$ versus $g$ in the first resonance interval for $N = 1$ is different from the first resonance interval for $N > 1$. This is because for $N = 1$ the L orbit which alone exists for a wide range of values of $g$ encloses  all the (two) equilibrium points  while for $N > 1$ the L orbit(s) encloses always three equilibrium points, though the number of equilibrium points is $N+1$. 

\begin{figure}[t]
\begin{center}
\epsfig{figure=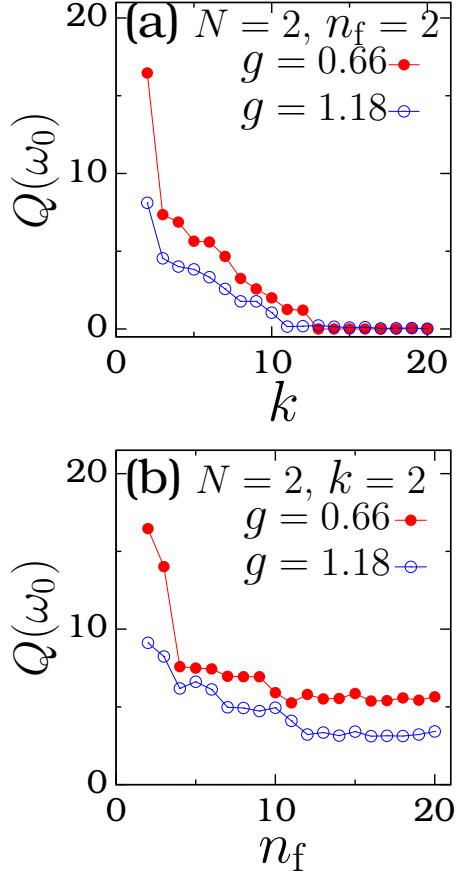, width=0.4\columnwidth}
\end{center}
\caption{Plots of (a) $Q(\omega_0)$ versus $k$ for $n_{\mathrm{f}} = 2$ and (b) $Q(\omega_0)$ versus $n_{\mathrm{f}}$ for $k = 2$ in $S(t)$ given by Eq.~(\ref{eq1}) and for $g = 0.66$ and $1.18$ in the first and second resonance intervals. At these two values of $g$ the response amplitude, $Q(\omega_0)$ becomes maximum for $N = 2$, $n_{\mathrm{f}} = 2$ and $k = 2$. }
\label{fig12}
\end{figure}

Finally, we study the effect of the number of periodic forces $n_{\mathrm{f}}$ and the value of $k$ on $Q(\omega_0)$. For $N = 2$, $n_{\mathrm{f}} = 2$ and $k = 2$, $Q(\omega_0)$ is maximum at $g = 0.66$ (in the first resonance interval) and $g = 1.18$ (in the second resonance interval). For these two fixed values of $g$, we compute $Q(\omega_0)$ for $k = 2 , 3, \cdots, 20$ with $n_{\mathrm{f}}=2$ and for $n_{\mathrm{f}} = 2, 3, \cdots, 20$ with $k = 2$. The result is presented in Fig.~\ref{fig12}. For fixed values of $n_{\mathrm{f}}$, $Q(\omega_0)$ decays to zero when $k$ increases. But for $k = 2$ when $n_{\mathrm{f}}$ increases, $Q(\omega_0)$ decreases and then becomes a nonzero constant value. The above result shows that the choice $n_{\mathrm{f}} = k = 2$ is a better choice compared to other choices of $n_{\mathrm{f}}$ and $k$.

%
\section{Conclusion} 
 \label{concl}
Weak signal detection and its amplification is an important process in certain nonlinear systems. A weak signal can be amplified by a relatively high-frequency input signal. It is important to study the characteristic properties of nonlinear functions on the high-frequency induced resonance. In the present paper we have focused our analysis on the influence of the number of breakpoints on the high-frequency induced resonance in a modified Chua's circuit with (i) single low-frequency input signal and (ii) multiple low-frequency signal. In both cases resonance at the frequencies present in the input signal is observed. However, in the cases of multi-frequency force with the form considered in Eq.~(\ref{eq1}) resonance at the missing frequencies is also observed. An interesting feature of the Chua's  circuit is that the number of breakpoints in the characteristic function can be easily varied. We have shown that the number of resonances is equal to the number of breakpoints. Thus, the number of resonances can be easily varied by varying the number of breakpoints. Figure~\ref{fig8} shows that $Q(\omega_0)\approx 0$ for $g$ values outside the resonance intervals. Consequently, one can either maximize or suppress the fundamental missing frequency $\omega_0$ by appropriately choosing the value of $g$. Here, we have restricted our analysis to the role of the number of breakpoints. The study of the influence of barrier height at the breakpoints, the spacing between the successive breakpoints on the two types of resonance considered here provide further interesting results.

\subsection*{Acknowledgments}
KA acknowledges the support from University Grants Commission (UGC), India in the form of UGC-Rajiv Gandhi National Fellowship. Financial support from the Spanish Ministry of Science and Innovation under Project No. FIS2009-09898 is acknowledged by MAFS.
\vskip 10pt
\noindent{\bf{REFERENCES}}

\begin{description}
\item{}
Alaoui, M.A.A. [1999] ``Differential equations with multispiral attractors,'' {\emph{Int. J. Bifurcation and Chaos}} {\bf{9}}, 1009-1039.
\item{}
Arathi, S., Rajasekar, S. \& Kurths, J. [2013] ``Stochastic and coherence resonance in a modified Chua's circuit system with multi-scroll orbits,'' {\emph{Int. J. Bifurcation and Chaos}}, (2013, in press).
\item{}
Balenzuela, P., Braun, H. \& Chialvo, D.R., [2012] ``The ghost of stochastic resonance: an introductory review,'' {\it Contemporary Physics}, 17--38.
\item{}
Blekhman, I.I. \& Landa, P.S. [2004] ``Conjugate resonances and bifurcations in nonlinear systems under biharmonical excitation,'' {\emph{Int. J. Non-linear Mechanics}} {\bf{39}}, 421-426.
\item{}
Boukabou, A., Sayoud, B., Boumaiza, H. \& Mansouri, N. [2009] ``Control of $n$-scroll Chua's circuit,'' {\emph{Int. J. Bifurcation and Chaos}} {\bf{19}}, 3813-3822.
\item{}
Buld\'{u}, J.M., Chialvo, D.R., Mirasso, C.R., Torrent, M.C. \& Garc\'{i}a-Ojalvo, J. [2003] ``Ghost resonance in a semiconductor laser with optical feedback,'' {\emph{Europhys. Lett.}} {\bf{64}}, 178-184.
\item{}
Buld\'{u}, J.M., Gonz\'{a}lez, C.M., Trull, J., Torrent, M.C. \& Garc\'{i}a-Ojalvo, J. [2005] ``Coupling-mediated ghost resonance in mutually injected lasers,'' {\emph{Chaos}} {\bf{15}}, 013103.
\item{}
Calvo, O. \& Chialvo, D.R. [2006] ``Ghost stochastic resonance in an electronic circuit,'' {\emph{Int. J. Bifurcation and Chaos}} {\bf{16}}, 731-735.
\item{}
Campos-Canton, E., Barajas-Ramirez, J.G., Solis-Perales, G. \& Femat, R. [2010] ``Multiscroll attractors by switching systems,'' {\emph{Chaos}} {\bf{20}}, 1-6.
\item{}
Chialvo, D.R., Calvo, O., Gonzalez, D.L., Piro, O. \& Savino, G.V. [2002] ``Subharmonic stochastic synchronization and resonance in neuronal systems,'' {\emph{Phys. Rev. E}} {\bf{65}}, 050902.
\item{}
Chialvo, D.R.[2003] ``How we hear what is not there: A neural mechanism for the missing fundamental illusion,'' {\emph{Chaos}} {\bf{13}}, 1226-1230.
\item{}
Chizhevsky, V.N. \& Giacomelli, G. [2008] ``Vibrational resonance and the detection of aperiodic binary signals,'' {\emph{Phys. Rev. E}} {\bf{77}}, 051126.
\item{}
Deng, B., Wang, J., Wei, X., Tsang, K.M. \& Chan, W.L. [2010] ``Vibrational resonance in neuron populations,'' {\emph{Chaos}} {\bf{20}}, 013113.
\item{}
Fortuna, L., Frasca, M. \& Xibilia, M.G., [2009] {\emph{Chua's Circuit Implementations: Yesterday, Today and Tomorrow}} (World Scientific, Singapore).
\item{}
Gomes, I., Vermelho, M.V.D. \& Lyra, M.L. [2012] ``Ghost resonance in the chaotic Chua's circuit,'' {\emph{Phys. Rev. E}} {\bf{85}}, 056201.
\item{}
Jeyakumari, S., Chinnathambi, V., Rajasekar, S. \& Sanjuan, M.A.F. [2009] ``Single and multiple vibrational resonance in a quintic oscillator with monostable potentials,'' {\emph{Phys. Rev. E}} {\bf{80}}, 046608.
\item{}
Khibnik, A.I., Roose, D. \& Chua, L.O. [1993] ``On periodic orbits and homoclinic bifurcations in Chua's circuit with a smooth nonlinearity,'' {\emph{Int. J. Bifurcation and Chaos}} {\bf{3}}, 363-384.
\item{}
Kilic, R., [2010] {\emph{A Practical Guide For Studying Chua's Circuits}} (World Scientific, Singapore).
\item{}
Lamarque, C.-H., Janin, O. \& Awrejcewicz, J. [1999] ``Chua systems with discontinuities,'' {\emph{Int. J. Bifurcation and Chaos}} {\bf{9}}, 561-616.
\item{}
Landa, P.S. \& McClintock, P.V.E. [2000] ``Vibrational resonance,'' {\emph{J. Phys. A: Math. Gen.}} {\bf{33}}, L433-L438.
\item{}
Lopera, A., Buld\'{u}, J.M., Torrent, M.C., Chialvo, D.R. \& Garcia-Ojalvo, J. [2006] ``Ghost stochastic resonance with distributed inputs in pulse-coupled electronic neurons," {\emph{Phys. Rev. E}} {\bf{73}}, 021101.
\item{}
L\"{u}, J., Chen, G., Yu, X. \& Leung, H. [2004] ``Analysis and design of multi-scroll chaotic attractors from saturated function series,'' {\emph{IEEE Trans. Circuits Syst. I: Fund. Th. Appl.}} {\bf{51}}, 2476-2490.
\item{}
Lu, J. \& Chen, G. [2006] ``Generating multiscroll chaotic attractors: Theories, methods and applications," {\emph{Int. J. Bifurcation and Chaos}} {\bf{16}}, 775-858.
\item{}
Madan, R.N. [1993] {\emph{Chua's Circuit: A Paradigm for Chaos}} (World Scientific, Singapore).
\item{}
Mahla, A.I. \& Badan Palhares, \'{A}.G. [1993] ``Chua's circuit with a discontinuous nonlinearity,'' {\emph{J. Circuit Systems and Computers}} {\bf{3}}, 231-237.
\item{}
\"{O}zo\v{g}uz, S., Elwakil, A.S. \& Salama, K. [2002] ``$n$-scroll chaos generator using nonlinear transconductor,''  {\emph{Electronics Letters}} {\bf{38}}, 685-686.
\item{}
Rajasekar, S., Abirami, K. \& Sanju\'{a}n, M.A.F. [2011] ``Novel vibrational resonance in multistable systems,'' {\emph{Chaos}} {\bf{21}}, 033106.
\item{}
Sakthivel, G., Rajasekar, S., Thamilmaran, K. \& Dana, S.K. [2012] ``Statistical measures and diffusion dynamics in a modified Chua's  circuit equation with multiscroll attractors,'' {\emph{Int. J. Bifurcation and Chaos}} {\bf{22}}, 1250004.
\item{}
Suykens, J.A.K. \& Vandewalls, J. [1993] ``Generation of $n$-double scrolls $(n=1,\;2,\;3,\;4,\cdots)$,'' {\emph{IEEE Trans. Circuits Syst. I: Fund. Th. Appl.}} {\bf{40}}, 861-867.
\item{}
Ullner, E, Zaikin, A., Garc\'{i}a-Ojalvo, J., B\'{a}scones, R. \& Kurths, J. [2003] ``Vibrational resonance and vibrational propagation in excitable systems,'' {\emph{Phys. Lett. A}} {\bf{312}}, 348-354.
\item{}
Xia, X.H., Chen, G.R. \& Gai, R.D. [2003] ``On control Lyapunov modes of linear control systems,'' {\emph{Control Theory and Applications}} {\bf{20}}, 223-227.
\item{}
Xu, F., Yu, P. \& Liao, X. [2009] ``Global analysis on $n$-scroll chaotic attractors of modified Chua's circuit,'' {\emph{Int. J. Bifurcation and Chaos}} {\bf{19}}, 135-157.
\item{}
Xu, F. \& Yu, P. [2009] ``Global stabilization and synchronization of $N$-scroll chaotic attractors in a modified Chua's circuit with hyperbolic tangent function,'' {\emph{Int. J. Bifurcation and Chaos}} {\bf{19}}, 2563-2572.
\item{}
Yal\c{c}in, M.E., \"{O}zo\v{g}uz, S., Suykens, J.A.K., \& Vandewalle, J. [2001] ``$n$-scroll chaos generators: a simple circuit model,'' {\emph{Electronics Letters}} {\bf{37}}, 147-148. 
\item{}
Yu, S., L\"{u}, J., Leung, H. \& Chen, G. [2005] ``Design and implementation of $n$-scroll chaotic attractors from a general jerk circuit,'' {\emph{IEEE Trans. Circuits Syst. I: Fund. Th. Appl.}} {\bf{52}}, 1459-1476.
\item{}
Yu, S., Tang, W.K.S. \& Chen, G. [2007] ``Generation of $n\times m$-scroll attractors under a Chua-circuit framework,'' {\emph{Int. J. Bifurcation and Chaos}} {\bf{17}}, 3951-3964.
\item{}
Zhong, G.Q., Man, K.F. \& Chen, G. [2002] ``A systematic approach to generating $n$-scroll attractor,'' {\emph{Int. J. Bifurcation and Chaos}} {\bf{12}}, 2907-2915.
\item{}
Zou, Y.-L., Zhu, J. \& Chen, G. [2006] ``Adaptive control of chaotic $n$-scroll Chua's circuit,'' {\emph{Int. J. Bifurcation and Chaos}} {\bf{16}}, 1089-1096.
\end{description}

\end {document}